\documentclass[twocolumn,pre,showpacs,amsmath]{revtex4}
\usepackage[xdvi]{graphicx}
\usepackage{url}

\newcommand{\pder}[2]{\frac{\partial #1}{\partial  #2}}

\newcommand{\der}[2]{\frac{d #1}{d  #2}}

\newcommand{\ep}{\epsilon}
\newcommand{\eq}{\mathrm{eq}}
\newcommand{\rmG}{\mathrm{G}}
\newcommand{\rmL}{\mathrm{L}}
\newcommand{\rmGL}{\mathrm{L/G}}

\newcommand{\pex}{p^\mathrm{ex}}
\newcommand{\bT}{\tilde T}
\newcommand{\mT}{T_\mathrm{m}}
\newcommand{\Tc}{T_\mathrm{c}}
\newcommand{\xc}{X}

\begin{document}
\title{Liquid-gas transitions in steady heat conduction}

\author{Naoko Nakagawa}
\affiliation {Department of Physics,  
              Ibaraki University, Mito 310-8512, Japan}

\author{Shin-ichi Sasa}
\affiliation {
Department of Physics, Kyoto University, Kyoto 606-8502, Japan}

\date{\today}

\begin{abstract}
We study liquid-gas transitions 
of heat conduction systems in contact with two heat baths
under constant pressure in the linear response regime.
On the basis of local equilibrium thermodynamics,
we propose an equality with a global temperature, which determines
the volume  near the equilibrium liquid-gas transition.
We find that the formation of the liquid-gas interface is accompanied by 
a discontinuous change in the volume
when increasing the mean temperature of the baths.
A super-cooled gas  near the interface is
observed  as a stable steady state.
\end{abstract}

\maketitle

{\em Introduction.---}
Liquid-gas transitions under constant pressure have been a classical subject of equilibrium thermodynamics \cite{Callen}.
In reality, however, a temperature gradient is formed, and thus the transition properties may be influenced by heat flow. 
As related experiments, enhanced heat conduction by condensation and evaporation was observed in turbulent systems \cite{Zhong, Urban}.
In order to describe such nonequilibrium phenomena systematically, we first need to establish a thermodynamic theory for phase transitions under heat conduction.

\begin{figure}[b]
\begin{center}
\includegraphics[scale=0.253]{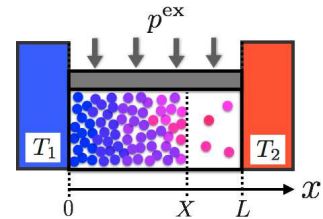}
\caption{Schematic illustration of experimental systems.}
\label{fig1}
\end{center}
\end{figure}

As the simplest situation, we consider cases where the pressure
and heat flux are spatially homogeneous, which is
illustrated in  Fig. \ref{fig1}.  
Specifically, let $\Tc(\pex)$ be the temperature
for the liquid-gas transition in equilibrium under constant pressure $\pex$.
When $\Tc(\pex)$ is between the temperatures of the baths \cite{fn:GLTC}, 
there is no reliable theory for determining
the steady state  even in the linear response regime.
Indeed, the standard hydrodynamic equations \cite{Landau} have many stationary solutions \cite{fn:hydro} once the liquid-gas interface is contained \cite{PA,vW,Bedeaux03,Onuki}. 
Furthermore, since the density profile has to be determined under the constraint of global mass conservation, the variational principle for selecting the steady state, if it exists, should be formulated for the whole system. Such a theory has not been reported yet.

   
  Over the last two decades, statistical mechanics of nonequilibrium systems has progressed significantly \cite{Sekimoto-book,SeifertRPP,information-thermodynamics}, 
  owing to the discovery of universal relations associated with the second law of thermodynamics
  \cite{Evans-Cohen-Morriss,Gallavotti,JarzynskiPRL,Kurchan,LS,Maes,Crooks}.
  As examples that may be related to the above problem, we point out 
  extensions of thermodynamic relations \cite{Hatano-Sasa,KNST, infinite-family, NN, Jona-thermo, Maes-thermo},
  variational formulas associated with large deviation theory
  \cite{Bodineau-Derrida,Maes-LD, Nemoto,efficiency,Bertini-rev},
  representations of steady state probability densities \cite{KN,KNST-rep,
    Maes-rep},
  and inequalities stronger than the second law \cite{BS,dissipation-bound,Shi-Sa-Ta,Edgar}.
  However, these results are not directly applicable to the analysis
  of liquid-gas transitions in heat conduction. 

In this Letter, we generalize an equilibrium variational principle that determines the volume near the liquid-gas transition. Concretely, on the basis of local equilibrium thermodynamics in the linear response regime, we propose the equality (\ref{main}) with a global temperature $\bT$, the main claim of this Letter, which corresponds to the generalized variational principle. This allows us to obtain the phase diagram of the heat conduction system, which can be examined in experiments.

{\em Setup.---}
We study the system shown in Fig. \ref{fig1}.
A heat bath of temperature $T_1$ is attached to the left
end $(x=0)$ of the system, and a second heat bath of temperature $T_2$
to the right end $(x=L)$, where $T_1 \le T_2$ is assumed
without loss of generality.
We focus on cases that $\Tc(\pex)$ is far below the liquid-gas critical temperature.
The length $L$ of the system is fixed throughout this Letter.
Other boundaries are  thermally insulating. The top plate
is freely movable under constant pressure $\pex$.  
For simplicity, gravity effects are ignored.


We first consider the equilibrium case $T_1=T_2=T$. Let $V$ and $N$ be the volume of
the system and the number of particles in moles. 
As an example, we take the van der Waals equation of state
\begin{equation}
  p=\frac{RT \rho}{1- b \rho}-a \rho^2
\label{van-der}
\end{equation}  
and the heat capacity $C_{V}=\eta NR$, where $R$ is the gas constant,
$a$, $b$ and $\eta$ are constants depending on the material, and 
$\rho=N/V$. 
Note that (\ref{van-der}) represents even meta-stable
states. 
The van der Waals free energy $F_{\rm vW}(T,V)$ \cite{fn:vW}
defined by $p= -\partial F_{\rm vW}(T,V)/\partial V$
is derived as
\begin{equation}
-N RT \log \left[T^{\eta}\left(\frac{V}{N}-b \right)\right]
-a \frac{N^2}{V}+(c_1T +c_2) N,
\label{free}
\end{equation}  
where  $c_1$ and $c_2$ are arbitrary constants that depend on
the reference state for the entropy and the energy, respectively. 
Let $\rho^\rmL$ and $\rho^\rmG$  be densities 
corresponding to liquid state and gas state. The two densities
satisfy $p(T,\rho^\rmL)=p(T,\rho^\rmG)=\pex$
with $\rho^\rmL >\rho^\rmG$. We then express the thermodynamic value
of $V=N/\rho$, which is either $V=N/\rho^\rmL$ or $V=N/\rho^\rmG$, 
by $V_*(T,\pex)$. 
For the following  variational function with $(T,\pex)$ fixed:
\begin{equation}
 {\cal G}_\eq (V;T,\pex) \equiv  F_{\rm vW}(T,V)+\pex V,
\label{eq:gibbs}
\end{equation}
$V_*(T,\pex)$ is characterized by  the variational
principle ${\cal G}_\eq(V_*(T,\pex);T,\pex) \le {\cal G}_\eq(V; T, \pex)$
for any $V$. 
There exists $\Tc(\pex)$ at which $ V_*(T,\pex)$
is discontinuous as a function of $T$ \cite{fn:Legendre}.
This singular behavior corresponds to the liquid-gas
transition in equilibrium systems, and it is described by the thermodynamic Gibbs
free energy  $G(T,\pex)\equiv {\cal G}_\eq (V_*(T, \pex) ;T,\pex)$.
 For hard spheres
with long-range attractive interaction,
 $\Tc(\pex)$ is exactly determined by means of the variational principle
 with (\ref{free}) and (\ref{eq:gibbs})
\cite{KUH,Kampen,fn:gen}.


{\em Main result.---}
We consider steady heat conduction states. 
We set $\Delta \equiv T_2-T_1 >0 $ and $\ep \equiv \Delta/T_1$.
We focus on the linear response regime where $\ep \ll 1$.
Since gravity effects are ignored,
the heat conduction state is  homogeneous in directions perpendicular to $x$ \cite{fn:bubbles}. 
We choose the mean temperature
$\mT\equiv (T_1+T_2)/2$ as a control parameter. Let $\xc$ be the position
of the interface 
between the liquid region $0 \le x < \xc$ and the gas region
$ \xc < x \le L$. That is, for a given $T(x)$, we set
$\rho(x)=\rho^\rmL(x)$ in $x < X$ and $\rho(x)=\rho^\rmG(x)$ in $x > X$,
where the pressure balance equation
\begin{equation}
p(T(x),\rho(x)) = \pex
\label{p-balance}  
\end{equation}  
holds. Note that $\rho(x)$ is discontinuous only at the interface $x=\xc$.
The continuous temperature profile is determined by the conductivity
$\kappa(T,\rho)$. Explicitly, $T(x)$ satisfies 
\begin{equation}
  -\kappa(T(x),\rho(x)) \partial_x T =J,
  \label{Fourier}
\end{equation}  
where $J$ is constant in $x$, and $T(0)=T_1$ and $T(L)=T_2$.
In this manner, $T(x)$ and $\rho(x)$ are determined from
(\ref{p-balance}) and (\ref{Fourier}) for a given $X$.
Since the volume $V$ of the system is obtained by
\begin{equation}
  \frac{V}{L} \int_0^L dx~ \rho(x) = N,
\label{v-det}
\end{equation}
$V$ has one-to-one correspondence with $X$. Thus, the
solutions $T(x)$ and $\rho(x)$ satisfying (\ref{p-balance})
and (\ref{Fourier}) may be parametrized by $V$.
We express the solutions and the interface position as $T(x;V)$, $\rho(x;V)$ and $X(V)$, respectively. 
For the steady state value $V_*$, we set $X_*=X(V_*)$. 
Furthermore, we define $X_*=0$ or $X_*=L$ when the space is filled with either gas or liquid, respectively. We next propose a formula for determining $V_*$.

Since local thermodynamic quantities characterize the steady 
heat conduction state in the linear response regime, a candidate
for the variational function is
\begin{equation}
\frac{V}{L} \int_0^L dx~ [f(T(x;V),\rho(x;V))+\pex],
\label{vf}
\end{equation}
which is the natural extension of the right-hand side in
(\ref{eq:gibbs}). Here, $f(T,\rho)=F_{\rm vW}(T,V)/V$ and
we ignore the free energy in the liquid-gas interface. 
Then, in order to identify fixed parameters, we use the fact
that $V_*$ has to be independent of $c_1$ and $c_2$ in (\ref{free}). 
Since the variational equation should be independent of $c_1$ and $c_2$,
we impose that  
\begin{equation}
\frac{V}{L} \int_0^L dx~ [ c_1 T(x;V)\rho(x;V) +c_2 \rho(x;V)]
\label{aribitrariness}
\end{equation}    
is kept constant with respect to the variation
in $V$. This means that $ V \int_0^L dx~ T(x;V)\rho(x;V)/L $
is a fixed parameter. Since this is proportional to
the temperature averaged over all  particles, 
we define a global temperature
\begin{equation}
\bT= \frac{\int_0^L dx~ T(x;V)\rho(x;V)}{\int_0^L dx~ \rho(x;V)}.
\label{global-T}
\end{equation}
The integral (\ref{aribitrariness}) is expressed 
by $(c_1\bT+c_2)N$ whose form  is the same as 
the last two terms of (\ref{free}). 
Thus, the variational function is expressed as
\begin{equation}
 {\cal G}(V;\bT,\pex, \Delta) \equiv \frac{V}{L}
 \int_0^L dx~ [f(T(x;V),\rho(x;V))
  +\pex]
     \label{vf-G}
\end{equation}
with $(\bT, \pex, \Delta)$ fixed.


The main claim of this Letter is that the equality
\begin{equation}
  \left.  \pder{ {\cal G}(V;\bT,\pex, \Delta)}{V} \right|_{V=V_*}=O(\ep^2)
\label{main}
\end{equation}  
holds for the steady state value $V_*$.
Here, $V_*$ is assumed to satisfy a scaling relation
that the thermodynamic state in the liquid or the gas region
persists for $(N, \Delta) \to (\lambda N, \lambda \Delta)$
with $1 \ll \lambda \ll \epsilon^{-1}$ 
\cite{fn:ss-ch}.
The derivation of  (\ref{main}) is given in the paragraphs including
(\ref{Phi-def-main}) and (\ref{t-gl}).
Using (\ref{main}), we can determine  $V_{*}(\bT,\pex)$
and $X_*$ as follows. 
First, we plot ${\cal G}(V; \tilde T, \pex, \Delta)$
as a function of $X=X(V)$ \cite{fn:gT}. When this graph shows a
local minimum at $X=X_*$ in $ 0 < X < L$, we find
the interface at $x=X_*$ because (\ref{main}) is satisfied.
When there is no local minimum,  $X_*$ is determined as 
either $X_*=0$ or $X_*=L$ which minimizes ${\cal G}$. 
Note that, for equilibrium cases $T_1=T_2=T$, 
there is no local minimum  when $T\neq\Tc(\pex)$.
The slope of ${\cal G}$ 
as a function of $X$ changes its sign at $T= \Tc(\pex)$ \cite{fn:LM}.
It is then found that $X_*=L$ for $T<\Tc(\pex)$ and $X_*=0$ for $T>\Tc(\pex)$.

\begin{figure}[tb]
\begin{picture}(180,240)(10,0)
\put(105,115){{\large$\bT$} ~(K)}
\put(-5,170){\rotatebox{90}{\large$\xc_*/L$}}
\put(-10,220){\Large (a)}
\put(96,157){\includegraphics[scale=0.38]{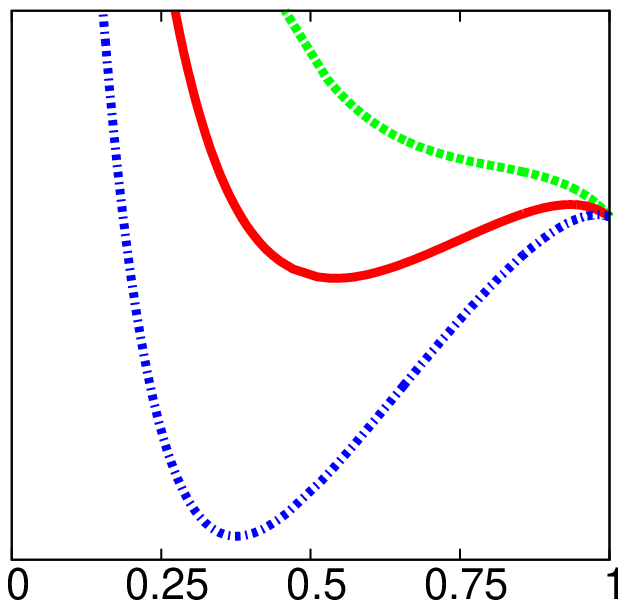}}
\put(144,159){{\tiny$\xc/L$}}
\put(110,177){\rotatebox{90}{\tiny${\cal G}(V; \bT,\pex,\Delta)$}}
\put(160,220){{\tiny$258$K}}
\put(146,194){{\tiny$259$K}}
\put(160,180){{\tiny$260$K}}
\put(-10,115){\includegraphics[scale=0.70]{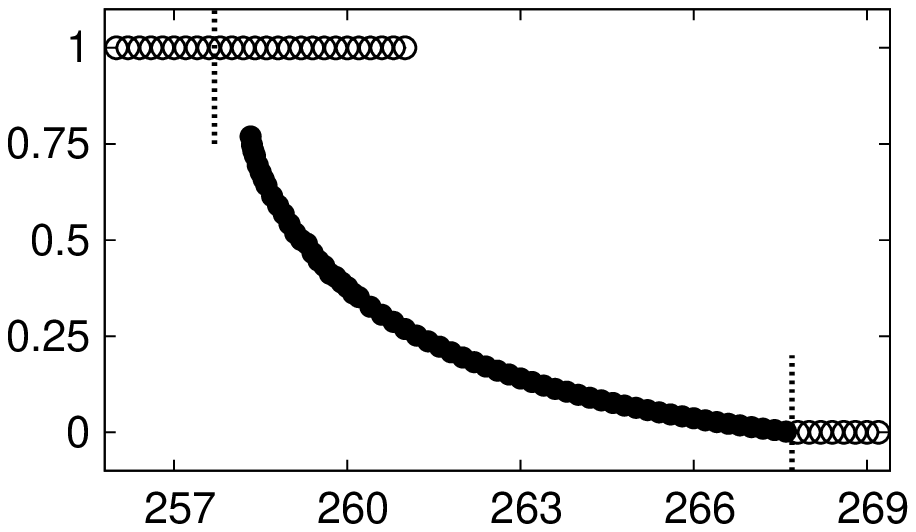}}
\put(105,0){{\large$\mT$} ~(K)}
\put(-5,50){\rotatebox{90}{\large$\xc_*/L$}}
\put(-10,95){\Large (b)}
\put(140,92){\scriptsize{interface}}
\put(140,80.){\scriptsize{no\! interface}}
\put(30,21){\includegraphics[scale=0.245]{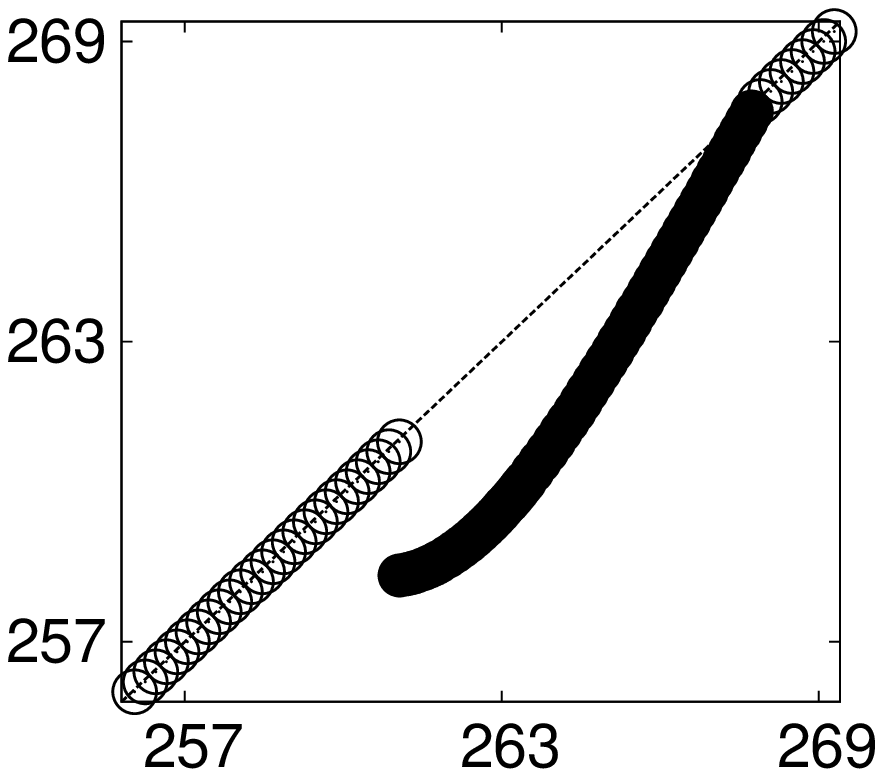}}
\put(38,81){\scriptsize$\bT$}
\put(97,28){\scriptsize$\mT$}
\put(-10,-2){\includegraphics[scale=0.70]{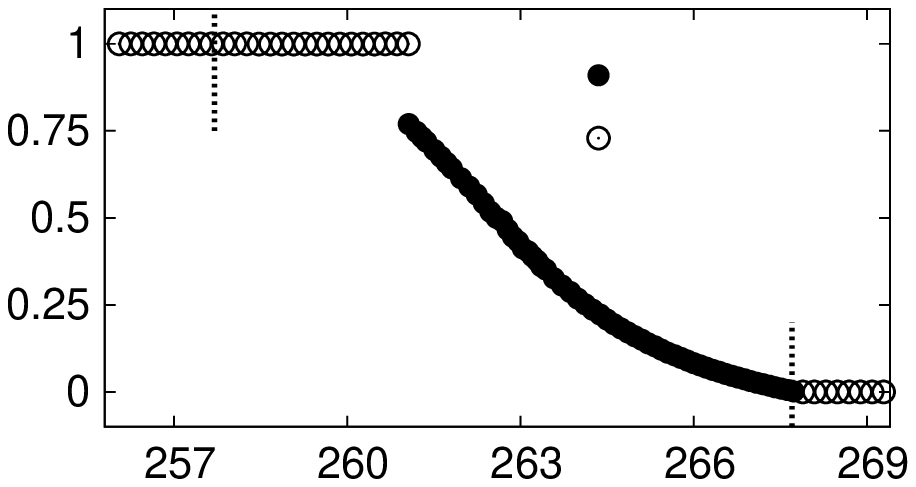}}
\end{picture}
\caption{(Color online) Interface position $\xc_*$ (filled circles) as a function of (a) $\bT$ and (b) $\mT$.
  Open circles represent $\xc_*=L$ and $\xc_*=0$.
  Dotted lines are $\mT=257.7$K (corresponding to $T_2=\Tc(\pex)$)
  and $267.7$K (corresponding to $T_1=\Tc(\pex)$),
where $\Tc(\pex)=262.7$ K.
   The inset of (a) shows examples of the variational function with
  local minimum in $0<X<L$ for higher $\bT$.
    The inset of (b) shows a graph of the map
  from $\mT$ to $\bT $.
  The parameter values are
  $a=0.365$ Pa$\cdot$ m$^6$/mol and $b=4.28\times 10^{-5}$ m$^3$/mol
of the van der Waals equation for ${\rm CO}_2$ \protect{\cite{CO2}}. 
 $C_V=5NR$, where $R=8.31$ J/K$\cdot$mol,
and $\kappa=0.1$ W/m$\cdot$ K in the liquid branch $\rho > \rho_c$ and $\kappa=0.02$ W/m$\cdot$ K
in the gas branch $\rho < \rho_c$, where $\rho_c=10^4$ mol/m$^3$
referring to the database \protect{\cite{NIST}}.
 $N=1$ mol without loss of generality, $\pex=4.0\times 10^6$ Pa and $\Delta =10$ K.
   }
\label{fig2}
\end{figure}

\begin{figure}[tb]
\begin{picture}(180,220)(10,0)
\put(95,0){\large $x/L$}
\put(0,45){\rotatebox{90}{{\large$\mu$}~~{\footnotesize(J/mol)}}}
\put(0,115){\rotatebox{90}{{\large $\rho$}~\footnotesize (mol/m$^3$)}}
\put(0,180){\rotatebox{90}{\large $T$~{\footnotesize (K)}}}
\put(168,190){$\large T_\mathrm{c}(\pex)$}
\put(175,147){\scriptsize gas}
\put(175,137){\scriptsize liquid}
\put(175,127){\scriptsize liquid-gas}
\put(0,146.2){\includegraphics[scale=0.55]{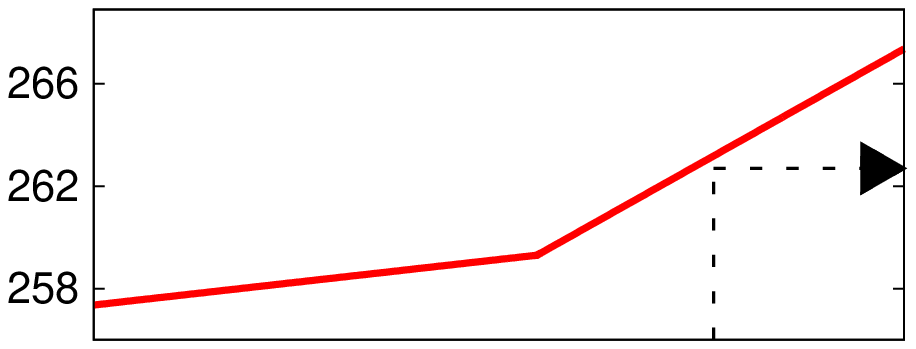}}
\put(0,94){\includegraphics[scale=0.55]{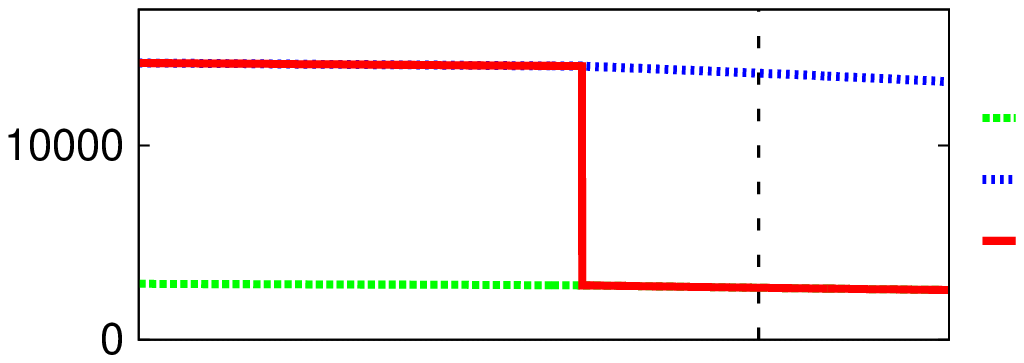}}
\put(0,0){\includegraphics[scale=0.55]{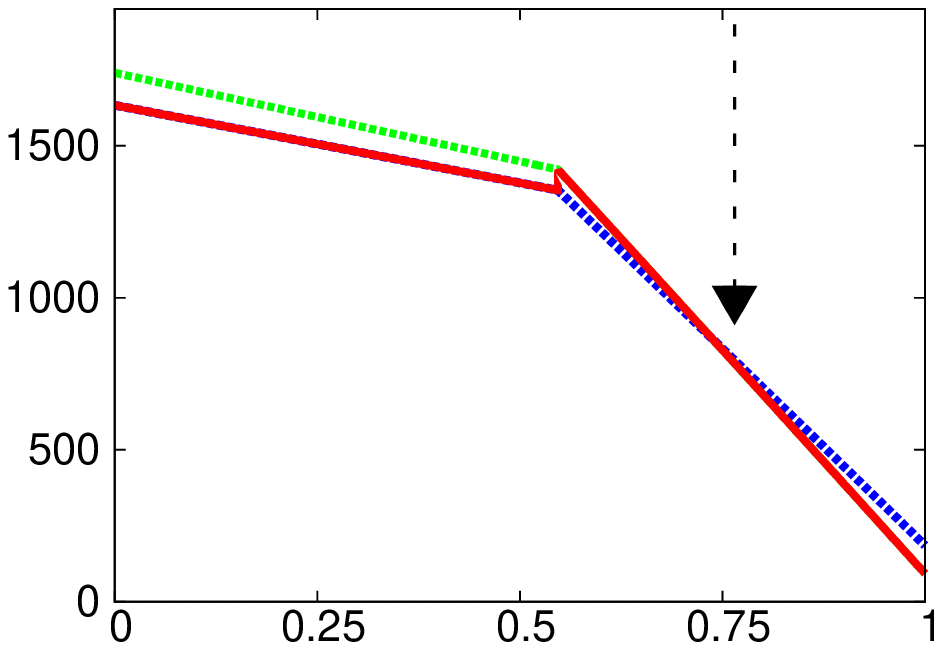}}
\put(32,17){\includegraphics[scale=0.25]{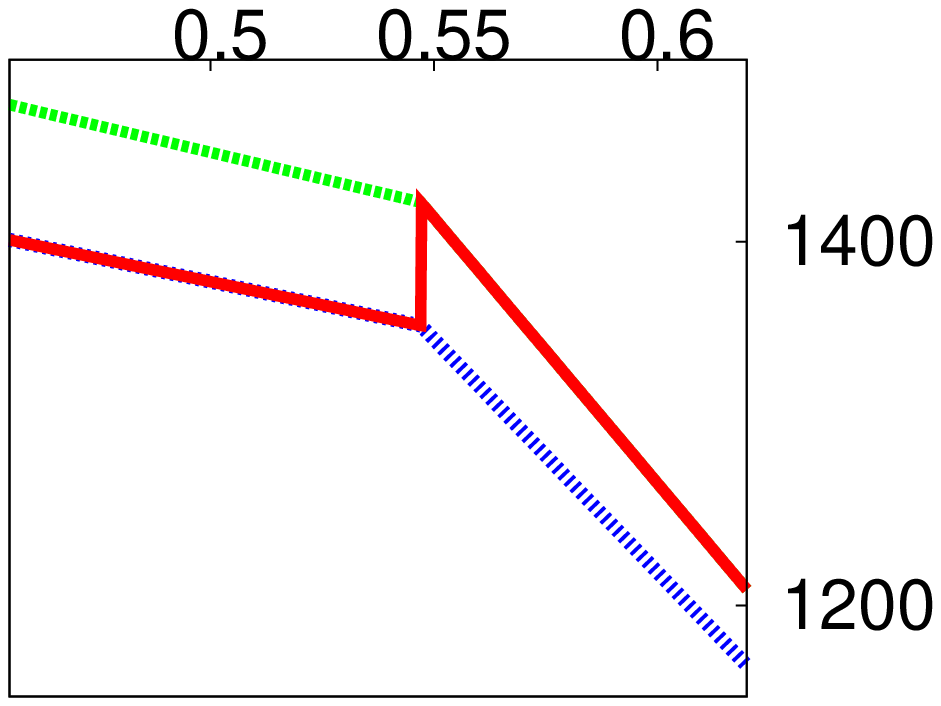}}
\end{picture}
\caption{Profiles of local thermodynamic variables for $\bT=259.0$ K and
  $T_2-T_1=10.0$ K ($T_1=257.36$ K, $T_2=267.36$ K).  The temperature $T(x;V_*)$, 
   the density $\rho(x;V_*)$ 
  and the chemical potential  $\mu(x;V_*)=\mu(T(x;V_*),\rho(x;V_*))$
     are shown   with red lines
in the top, middle and bottom panels, respectively.
 Green lines correspond to
  $\rho^\rmG(x)$ and $\mu(T(x;V_*),\rho^\rmG(x))$ in the middle and bottom panels,
  and blue lines correspond to $\rho^\rmL(x)$ and 
  $\mu(T(x;V_*),\rho^\rmL(x))$. The inset of the bottom panel is
  the close-up view of the chemical potential profiles around the interface.
  The liquid-gas interface exists at $\xc_*/L=0.547$. 
  The constants $c_1$ and $c_2$ are chosen as $c_1=-113.45$ J/mol$\cdot$ K 
  and $c_2=7400$ J/mol.}
\label{fig4}
\end{figure}

{\em Example.---}  
Fig.~\ref{fig2}(a) shows the graph of the interface position
$\xc_{*}$ as a function of $\bT$ for the system with $\Tc(\pex)=262.7$ K.
Since  $\bT$ is not an experimentally controllable parameter,
we employ $\mT$ so as to predict phenomena in experiments. 
The relation between $\bT$ and $\mT$ is shown in the inset of Fig.~\ref{fig2}(b).
When there is no interface, $\tilde T=\mT+O(\ep^2)$ holds \cite{fn:Tm}. 
By using the relation in the inset, 
we draw a graph of $\xc_*$ as a function of $\mT$  in Fig.~\ref{fig2}(b).
We find that  the transition 
from $\xc_*=L$ to $0 < \xc_* < L$ is discontinuous.
Since $T_2>\Tc(\pex)$ on the right side of the left dotted line in Figs.~\ref{fig2},
the whole system is filled with liquid 
even when $T_2$ is slightly larger than $\Tc (\pex )$.
This means that the super-heated liquid is stable
near the right boundary.  On the other hand, as $T_1-\Tc(\pex) \to 0$,
which is indicated by the right dotted line in Figs.~\ref{fig2}, 
the liquid region disappears  continuously.


The discontinuous transition is connected to the standard liquid-gas
transition when $\Delta \to 0$. 
However, the nature of the discontinuous transition is rather different. First, 
if the local temperature of the interface were always identical
to the equilibrium transition temperature $T_\mathrm{c}(\pex)$,  
$\xc_*$ would change continuously. Thus, the discontinuous transition
implies  that $T(\xc_*)-T_\mathrm{c}(\pex)\not = 0$,
which is indeed observed in the top panel of Fig. \ref{fig4}.
For the chemical potential $\mu(T,\rho)\equiv [f(T,\rho)+p(T,\rho)]\rho^{-1}$,
we plot its profile $\mu(T(x;V_*),\rho(x;V_*))$ as a function of $x$
 in the bottom panel of Fig. \ref{fig4}.  We then find the discontinuous
jump at the interface $x=\xc_*$.
This means $T(\xc_*)-T_\mathrm{c}(\pex)\not = 0$, because
$T_\mathrm{c}(\pex)$ is characterized by
$\mu(T_\mathrm{c}(\pex),\rho^\rmG)=\mu(T_\mathrm{c}(\pex),\rho^\rmL)$.
The position $\tilde x$ satisfying $T(\tilde x)=T_\mathrm{c}(\pex)$
is obtained from the crossing point of the two curves
$\mu(T(x;V_*),\rho^\rmG(x))$ and $\mu(T(x;V_*),\rho^\rmL(x))$,
as shown in Fig. \ref{fig4}. It should be noted that
in the region $ \xc_* < x < \tilde x$,
the super-cooled gas is observed as a stable steady state.


{\em Outline of the derivation of (\ref{main}).---} 
There are two key steps in the derivation of (\ref{main}).
The first step is that 
when there is no singularity of $\rho(x;V)$ in the region $I=[x_1,x_2]$,
the integration of a local quantity
$\phi(T(x),\rho(x))$ over the region $I$ is
estimated as 
\begin{equation}
\int_{x_1}^{x_2}  dx~ \phi(T(x;V),\rho(x;V))
=  |I| \phi(\mT^I,\bar \rho^I)+O(\ep^2),
\label{Phi-def-main}
\end{equation}
which follows from the trapezoidal rule for the integral
after the change of the integration variable from $x$ to $T$. 
Here, $\mT^I\equiv (T(x_1;V)+T(x_2;V))/2$,
$\bar \rho^I\equiv \int_{x_1}^{x_2}  dx~ \rho(x;V)/|I|$
and $|I|=x_2-x_1$.
The relation (\ref{Phi-def-main}) means that
a non-uniform system with $(T(x;V), \rho(x;V))$
is equivalent to an equilibrium system with  $(\mT^I, \bar \rho^I)$. 
We employ (\ref{Phi-def-main}) with $\phi=f$ or $\phi=\rho T$.

Next, we consider the case that the density is discontinuous
at $x=\xc$.  Since there is no singularity in the liquid region
$0 \le x < \xc$ or the gas region $\xc < x \le L$, we apply
(\ref{Phi-def-main}) to each region.
We put  $\rmL$ and $\rmG$ as the superscript on quantities,
respectively. 
By letting $N^\rmGL_*$  and $T^\rmGL_{\mathrm{m}*}$ as the steady state values,
it is assumed that the thermodynamic state in the liquid region, 
$(\pex, T^\rmL_{\mathrm{m}*}, N^\rmL_*)$, can be invariant under the scale transformations 
$(\Delta,N) \to ( \lambda \Delta, \lambda N)$
with $1 \ll \lambda \ll \epsilon^{-1}$, which corresponds to the 
extension of the gas region \cite{fn:inv-2}. This scaling assumption is expressed as
$\mT^\rmL(\pex, N^\rmL_*, \lambda \Delta,  \lambda N)
=\mT^\rmL(\pex, N^\rmL_*, \Delta,  N)=T^\rmL_{\mathrm{m}*}$.
Similarly, the scaling relation for keeping the thermodynamic state in
the gas region is also assumed. From these relations and
$ T^\rmG_{\mathrm{m}*} - T^\rmL_{\mathrm{m}*} =\Delta/2$, 
we obtain 
\begin{equation}
 T^\rmGL_{\mathrm{m}*} 
=\Tc(\pex) \mp \frac{\Delta}{2}\frac{N^\rmGL_*}{N}+O(\ep^2).
\label{t-gl}
\end{equation}  
This is the second key step in the derivation of (\ref{main}).

To evaluate the left-hand side of (\ref{main}),
we consider $ {\cal G}(V;\bT,\pex,\Delta)={\cal G}^{\rm L}+ {\cal G}^{\rm G}$.
We estimate $ {\cal G}^{\rm L}$ and ${\cal G}^{\rm G}$ using
(\ref{Phi-def-main}), and take the variation $V \to V+\delta V$ in
$ {\cal G}$ by fixing $\tilde T$, $\pex$ and $\Delta$.  The variation
$\delta V$ induces $\delta N^\rmG$,  $\delta N^\rmL$, $\delta \mT^\rmG$,
$\delta \mT^\rmL$, $\delta V^\rmG$ and  $\delta V^\rmL$.
The straightforward calculation using (\ref{t-gl}) leads to 
$ {\cal G}(V_*+\delta V;\bT,\pex, \Delta)-{\cal G}(V_*;\bT,\pex,\Delta)
=O(\ep^2)$.
This ends the proof of (\ref{main}) \cite{fn:proof}. 

\begin{figure}[bt]
\begin{picture}(150,117)(0,0)
\put(-5,15){\includegraphics[scale=0.3]{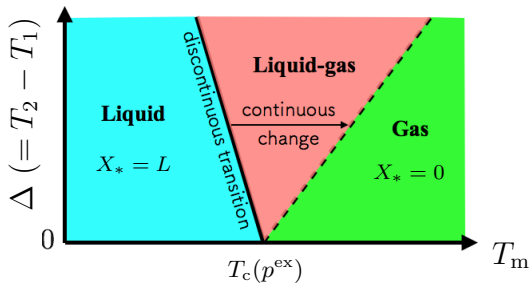}}
\put(-18,30){\rotatebox{90}{\large$\Delta ~(= T_2-T_1)$}}
\put(15,45){\footnotesize$\xc_*=L$}
\put(120,42){\footnotesize$\xc_*=0$}
\put(65,5){$\Tc(\pex)$}
\put(165,10){\large$\mT$}
\end{picture}
\caption{Schematic phase diagram in the heat conduction system
under constant pressure.}
\label{fig5}
\end{figure}

{\em Concluding Remarks.---} The result of our theory
is schematically summarized as the phase diagram in Fig. \ref{fig5}. 
We emphasize that either the super-cooled gas 
or super-heated liquid becomes stable as a local
equilibrium state in heat conduction. This striking phenomenon
is a consequence of the discontinuous transition from the liquid
to the liquid-gas coexistence state. Even without quantitative
measurements, observing this qualitatively new phenomenon in
experiments and numerical simulations would be very stimulating. 
Before ending this Letter,
we present a few remarks.

As a different approach to determine the position of the liquid-gas interface, the density-gradient dependent pressure 
\begin{equation}
\frac{1}{2}d_1\xi^2(\partial_x \rho)^2
  -d_2 \xi^2 \partial_x^2 \rho  +d_3 J \xi \partial_x \rho
\label{int-pressure}
\end{equation}
may be added to the left-hand side of (\ref{p-balance}), where $\xi$ is the width of the interface. For equilibrium cases $J = 0$, $d_1=-\rho^2 \partial (\rho^{-2}d_2)/\partial \rho $ is derived according to the van der Waals theory \cite{vW}. When this relation is applied to (\ref{int-pressure}) for the heat conduction states, the interface temperature turns out to deviate from $T_{\rm c}(\pex)$ by the influence of the $d_3$ term \cite{fn:SM1}. 
While the density-gradient terms (\ref{int-pressure}) are required to describe the density profile inside the interface, the variational principle (\ref{main}) determines the profile outside the interface. When the density profile inside the interface is not our concern, a density-gradient term  is not necessary in the variational functional (\ref{vf-G}). 
It should be noted that for equilibrium cases the density-gradient dependent pressure is systematically obtained from the free energy functional with the density-gradient term. It is an important future subject to have such a unified theory for heat conduction states.

Next, we remark on future theoretical studies. Since we focus on the linear response regime, we may use representations of the probability distribution and the variational principles for the steady state \cite{KN,KNST-rep,Maes-rep, Zubarev,Mclennan,Maes-SP,Maes-NJP}. 
  It is a challenging problem to derive (\ref{main}) on the basis of these results. 
  Related to this problem, one may study more general experimental configurations than Fig.~\ref{fig1}. 
  The variational function $\cal G$ can be similarly defined for such cases, but the variational problem cannot be easily solved because of the complicated geometry of the interface. 
  Another possible study is to seek an extended form of thermodynamics. 
  The liquid-gas coexistence predicted in this Letter may be interpreted as a phase separated by a first-order transition. We conjecture that the transition is characterized by the singularity of the generalized free energy $G(\bT,\pex,\Delta)\equiv {\cal G}(V_*(\tilde T,\pex,\Delta);\bT,\pex,\Delta) $
  that satisfies the fundamental relation in thermodynamics with the global temperature  $\tilde T$ \cite{NS}. 
  This framework is different from previous theories \cite{Keizer,Eu,Jou,Oono-paniconi,Sasa-Tasaki}. 
  When constructing generalized thermodynamics, we should carefully study the manner of contact \cite{Bertin,Seifert-contact,Dickman,Chiba-Nakagawa}. 
  Last but not least, we wish to extend our theory to describe thermodynamic phases of active matter \cite{Dhont, Solon} and phase transitions in turbulent flow \cite{Zhong,Urban}. 
  Although our theory is firmly formulated in the linear response regime, a framework using global quantities, which is our key concept, may be developed for the study of phenomena far from equilibrium. It would be quite interesting to discover new phenomena as the result of such a framework for general settings with various materials.

{\em Acknowledgment.---}  
The authors thank Kazuya Saito, Masato Itami, Christian Maes, 
Yohei Nakayama, Yoshi Oono, Hal Tasaki, Yuki Uematsu,
Yasuhiro Yamada, and Hiroshi Watanabe for their useful comments. They also thank Andreas 
Dechant for
critical reading of the manuscript.
The present study was supported by KAKENHI (Nos. 25103002, 15K05196, 17H01148). 



\widetext
\clearpage

\begin{center}
\textbf{\large Supplemental Material for \textit{
      Liquid-gas transitions in steady heat conduction}}\\
      
     \vspace{0.5cm}
{Naoko Nakagawa}\\
\textit{\small Department of Physics, Ibaraki University,
  Mito 310-8512, Japan}\\
     \vspace{0.2cm}  
{Shin-ichi Sasa}\\
\textit{\small Department of Physics, Kyoto University, Kyoto 606-8502, Japan}

\end{center}

\def\theequation{S\arabic{equation}}
\setcounter{equation}{0}


\section{Derivation of (11)}


We derive (11) in the main text.
In this section, we explicitly describe the dependence of $\cal G$ on $N$,
while we omit it in the main text.
For ${\cal G}(V;\bT,\pex,\Delta, N)$ defined by
\begin{equation}
  {\cal G}(V;\bT,\pex,\Delta, N) \equiv \frac{V}{L}
  \int_0^L dx [f(T(x;V),\rho(x;V))
    +\pex],
\label{G-def-supp}  
\end{equation}
we show 
\begin{equation}
  \left.  \pder{ {\cal G}(V;\bT,\pex, \Delta, N)}{V} \right|_{V=V_*}=O(\ep^2).
\label{main-supp}
\end{equation}  
In the above, $f(T,\rho)=F_{\rm vW}(T,V,N)/V$.


We first note the following {\it equivalence relation} 
for cases that there is no singularity of $\rho(x)$
in $I=[x_1,x_2]$. 
For a local quantity $\phi$, let $\Phi^I$ be 
\begin{equation}
\Phi^I \equiv  \frac{V}{|I|} \int_{x_1}^{x_2}  dx~ \phi(T(x;V),\rho(x;V)),
\label{Phi-def}
\end{equation}
where $|I|=x_2-x_1$. Then, $\Phi^I$ satisfies 
\begin{equation}
\Phi^I = V \phi(\mT^I, \bar \rho^I)+O(\ep^2),
\label{rule}
\end{equation}
where
\begin{eqnarray}
\mT^I  &=& \frac{T(x_1)+T(x_2)}{2},  \\
\bar \rho^I &=& \frac{1}{|I|}\int_{x_1}^{x_2} dx~ \rho(x).
\end{eqnarray}
This means that the space-integrated quantities  over the region $I$ 
for the nonuniform system with $(T(x;V),\rho(x;V))$ is equivalent to 
the thermodynamic quantities for the equilibrium system with
$(\mT^I, \bar \rho^I)$. 
The proof of (\ref{rule}) is given in Sec. \ref{lemma}.
In particular, by setting
$\phi=\rho(x;V)T(x;V)$, we obtain
\begin{equation}
  \tilde T=\mT+O(\ep^2)
\label{rule2}  
\end{equation}  
for systems without any singularity of $\rho(x;V)$
in $[0,L]$.


Now, we consider the heat conduction system in which
the density is discontinuous at $x=\xc$,
while there is no singularity in each
region  $0 \le x <\xc$ or $\xc < x \le L$.
We refer $0 \le x < \xc$ to as the liquid region 
and $\xc < x \le L$ to as the gas region, respectively.
We express quantities defined in these regions by putting the superscript  $\rmL$ or $\rmG$ on them.
Note that there are fixing conditions,
\begin{align}
&N^\rmL+N^\rmG=N,\\
&V^\rmL+V^\rmG=V,\\
&\bT^\rmL N^\rmL+\bT^\rmG N^\rmG=\bT N.\label{gTfix}
\end{align}
The last condition is due to the definition of $\bT$ given
in the main text as (9).
We then write
\begin{equation}
 {\cal G}={\cal G}^{\rmL}+{\cal G}^\rmG,
\end{equation}
where 
\begin{eqnarray}
  {\cal G}^\rmL  &=&   \frac{V}{L} \int_0^\xc  dx [f(T(x;V),\rho(x;V))+\pex], 
\label{Phi-def-L} \\
{\cal G}^\rmG  &=&   \frac{V}{L} \int_\xc^L dx [f(T(x;V),\rho(x;V))+\pex].
\label{Phi-def-G} 
\end{eqnarray}
In the present setting, the volumes $V^\rmL$ and $V^\rmG$ for the respective regions satisfy 
\begin{align}
V^\rmL/X=V^\rmG/(L-X)=V/L.
\end{align}
Therefore, each variational function is  expressed  consistently with 
the definition (\ref{G-def-supp}) as
\begin{align}
{\cal G}^\rmL&={\cal G}(V^\rmL;\bT^\rmL,\pex,\Delta^\rmL, N^\rmL)
=\frac{V^\rmL}{\xc} \int_0^\xc  dx [f(T(x;V),\rho(x;V))+\pex],\\
{\cal G}^\rmG&={\cal G}(V^\rmG;\bT^\rmG,\pex,\Delta^\rmG, N^\rmG)
=\frac{V^\rmG}{L-\xc} \int_\xc^L  dx [f(T(x;V),\rho(x;V))+\pex],
\end{align}
where $\Delta^\rmL=T(\xc)-T_1$ and $\Delta^\rmG=T_2-T(\xc)$.
Since there is no singularity in the respective regions, we can use the equivalence relation (\ref{rule}).
Then, we obtain
\begin{eqnarray}
 {\cal G}^\rmL &=&  V^\rmL [f(\mT^\rmL, \bar \rho^\rmL)+\pex] +O(\ep^2), \label{pre1} \\
 {\cal G}^\rmG &=&  V^\rmG [f(\mT^\rmG, \bar \rho^\rmG)+\pex] +O(\ep^2),\label{pre2}
\end{eqnarray}
where $\mT^\rmL=(T_1+T(\xc))/2$, $\mT^\rmG=(T_2+T(\xc))/2$,
$\bar \rho^\rmL =N^\rmL/V^\rmL$, and $\bar \rho^\rmG =N^\rmG/V^\rmG$. 

Recalling that $\mT^\rmGL=\bT^\rmGL+O(\ep^2)$, the fixing condition (\ref{gTfix}) leads to
\begin{equation}
\tilde T = \frac{\mT^\rmL  N^\rmL + \mT^\rmG  N^\rmG}{N}+O(\ep^2).
\label{tilde-t-supp}
\end{equation}
Then, by noting 
\begin{equation}
  \mT^\rmG-\mT^\rmL=\frac{\Delta}{2}, \label{t-g-l}
\end{equation}
we express $\mT^\rmL$ and $\mT^\rmG$ as
\begin{eqnarray}
\mT^\rmL =\tilde T-\frac{\Delta }{2}\frac{N^\rmG}{N}+O(\ep^2), \label{tmL} \\
\mT^\rmG =\tilde T+\frac{\Delta }{2}\frac{N^\rmL}{N}+O(\ep^2). \label{tmG}
\end{eqnarray}


We consider the variation $V \to V+\delta V$ for ${\cal G}$
with fixing $\tilde T$, $\pex$, and $\Delta$. The variation
$\delta V$ induces $\delta N^\rmG$,  $\delta N^\rmL$, $\delta \mT^\rmG$,
$\delta \mT^\rmL$, $\delta V^\rmG$,  and $\delta V^\rmL$,
where the fixing conditions lead to 
\begin{eqnarray}
  \delta N^\rmG&=& -\delta N^\rmL, \\
  \delta V^\rmG&=& -\delta V^\rmL +\delta V , \\
  \delta \mT^\rmG&=& \delta \mT^\rmL = \frac{\Delta}{2N} \delta N^\rmL.
\end{eqnarray}
By using (\ref{pre1}) and (\ref{pre2}), we obtain
\begin{eqnarray}
& &  {\cal G}(V+\delta V; \tilde T, \pex, \Delta,N)
 - {\cal G}(V; \tilde T, \pex, \Delta,N) \nonumber \\
&=&
F_{\rm vW}(\mT^\rmL+\delta \mT^\rmL,  V^\rmL+\delta V^\rmL, N^\rmL+\delta N^\rmL)
+
F_{\rm vW}(\mT^\rmG+\delta \mT^\rmG,  V^\rmG+\delta V^\rmG, N^\rmG+\delta N^\rmG)
 \nonumber \\
&-& F_{\rm vW}(\mT^\rmL,  V^\rmL, N^\rmL) - F_{\rm vW}(\mT^\rmG,  V^\rmG, N^\rmG)
+ \pex \delta V+O(\ep^2).
\label{deltaF}  
\end{eqnarray}
We define the chemical potential
\begin{equation}
 \mu(T,\rho)=\pder{F_{\rm vW}(T,V,N)}{N}
\label{mu-def}
\end{equation}  
with $\rho=V/N$, and the van der Waals entropy 
\begin{equation}
 S_{\rm vW}(T,V,N)=-\pder{F_{\rm vW}(T,V,N)}{T}.
\label{Sdef}
\end{equation}  
By noting the pressure balance 
\begin{equation}
  p(\mT^\rmL, \bar \rho^\rmL)=p(\mT^\rmG, \bar \rho^\rmG)=\pex,
\label{p-balance}
\end{equation}
(\ref{deltaF}) is transformed into
\begin{equation}
  {\cal G}(V+\delta V; \tilde T, \pex, \Delta,N)
 - {\cal G}(V; \tilde T, \pex, \Delta,N) 
= \left[
    -(S^\rmL+S^\rmG) \frac{\Delta}{2N}+\mu^\rmL-\mu^\rmG  \right] \delta N^\rmL +O(\ep^2),
\label{deltaF-2}  
\end{equation}
where
$S^\rmL\equiv S_{\rm vW}(\mT^\rmL,V^\rmL,N^\rmL)$,
$S^\rmG \equiv S_{\rm vW}(\mT^\rmG,V^\rmG,N^\rmG)$,
$\mu^\rmL\equiv \mu(\mT^\rmL,\bar \rho^\rmL)$
and $\mu^\rmG\equiv \mu(\mT^\rmG,\bar \rho^\rmG)$.


We next consider $\mu^\rmL-\mu^\rmG$
by interpreting $\rho$ to be a dependent variable of $(T,p)$,
$\rho=\rho(T,p)$. There exists the transition temperature $\Tc(\pex)$
for given $\pex$ such that  the density is discontinuous, i.e.,
\begin{equation}
\rho(\Tc(\pex)-0,\pex) \not = \rho(\Tc(\pex)+0,\pex),
\end{equation}  
whereas the chemical potential is continuous as
\begin{equation}
  \mu(\Tc(\pex), \rho(\Tc(\pex)-0,\pex))
=
 \mu(\Tc(\pex), \rho(\Tc(\pex)+0,\pex)).
\end{equation}
Using this equality, we express  $ \mu^\rmL-\mu^\rmG $ as
\begin{eqnarray}
  \mu^\rmL-\mu^\rmG
& =&
 \mu(\mT^\rmL, \rho(\mT^\rmL,\pex))-\mu(\Tc(\pex), \rho(\Tc(\pex)-0,\pex)) \nonumber \\
&- &
\mu(\mT^\rmG, \rho(\mT^\rmG,\pex))+\mu(\Tc(\pex), \rho(\Tc(\pex)+0,\pex)). 
\label{mu-dis-2}
\end{eqnarray}
We also note that the Gibbs-Duhem relation $N d\mu=-S_{\rm vW}dT + V d\pex$
leads to
\begin{equation}
  \frac{S_{\rm vW}(T,V)}{N} = -\pder{\mu(T,\rho(T,p))}{T}.
\label{GD}
\end{equation}
Here, since there is no singularity for respective regions
$0\le x< \xc$ and $\xc < x\le L$,  we can estimate
the first line and the  second line  in (\ref{mu-dis-2}) separately.
Substituting (\ref{GD}), we obtain 
\begin{equation}
\mu^\rmL-\mu^\rmG
= \frac{s(\mT^\rmL, \rho(\mT^\rmL,\pex))}{\bar \rho^\rmL}
(\Tc(\pex)-\mT^\rmL)
+
\frac{s(\mT^\rmG, \rho(\mT^\rmG,\pex))}{\bar \rho^\rmG}
(\mT^\rmG-\Tc(\pex))+O(\ep^2),
\end{equation}
with $s(T,\rho)=S_{\rm vW}(T,V,N)/V$. 


To this point, $V$ is not necessarily equal to $V_*$.
Then, for the steady state
value $V_*$, we express $\mT^\rmG{}_*$ and $\mT^\rmL{}_*$ as
\begin{eqnarray}
  \mT^\rmG{}_*  &=&  \Tc(\pex)+\frac{\Delta}{2}\frac{N^\rmG_*}{N}, \label{lam-1} \\
  \mT^\rmL{}_*  &=&  \Tc(\pex)- \frac{\Delta}{2}\frac{N^\rmL_*}{N} . \label{lam-2}
\end{eqnarray}
See Sec. \ref{lemma2} for the derivation of (\ref{lam-1}) and (\ref{lam-2}).
We then obtain 
\begin{eqnarray}
\mu^\rmL_*-\mu^\rmG_*
&=&  \frac{s(\mT^\rmL{}_*, \rho(\mT^\rmL{}_*,\pex))}{\bar \rho^\rmL_*}
         \frac{N^\rmL_*}{N} \frac{\Delta}{2}
+    \frac{s(\mT^\rmG{}_*, \rho(\mT^\rmG{}_*,\pex))}{\bar \rho^\rmG_*}
          \frac{N^\rmG_*}{N} \frac{\Delta}{2}  \nonumber \\
&=&  S_{\rm vW}(\mT^\rmL{}_*, V^\rmL_*, N^\rmL_*) \frac{\Delta}{2N}
      +S_{\rm vW}(\mT^\rmG{}_*, V^\rmG_*, N^\rmG_*) \frac{\Delta}{2N},
\end{eqnarray}
where we have used $\rho(T_\mathrm{m *}^\rmGL,\pex)=\bar\rho^\rmGL_*=N_*^\rmGL/V_*^\rmGL$.
By substituting this result into (\ref{deltaF-2}), we obtain
\begin{equation}
  {\cal G}(V_*+\delta V; \tilde T, \pex, \Delta,N)
 - {\cal G}(V_*; \tilde T, \pex, \Delta,N) = O(\ep^2).
\label{deltaF-3}  
\end{equation}
This leads to (\ref{main-supp}). 


\subsection{Proof of (\ref{rule})} \label{lemma}

In this section, we prove the equivalence relation (\ref{rule}).

We start with preliminaries. 
We take an interval $I=[x_1, x_2]$ in which there is no singularity
of $\rho(x;V)$.
For a smooth function $b(x)$, we consider the integral 
\begin{equation}
B\equiv \frac{1}{|I|}\int_{x_1}^{x_2} dx~ b(x),
\end{equation}  
where $|I|=x_2-x_1$.
Since $dT(x;V)/dx >0$, we transform the integration variable from
$x$ to $y=T(x;V)$. The function $x(y)$, which gives $x$ for $y$,
satisfies $y=T(x(y);V)$. We also introduce a function $J(y)$ by
\begin{equation}
\left.  \frac{dT(x;V)}{dx} \right|_{x=x(y)} =\frac{T(x_2)-T(x_1)}{|I|}J(y).
\end{equation}
We then have 
\begin{equation}
B=\frac{1}{T(x_2)-T(x_1)} \int_{T(x_1)}^{T(x_2)} dy~ \psi(y)
\label{B-def}
\end{equation}
with
\begin{equation}
  \psi(y)\equiv \frac{b(x(y))}{J(y)}.
\end{equation}
Here, we expand   $\psi(y)$ as
\begin{equation}
  \psi(y)=\psi(\mT^I)+ (y-\mT^I) \left. \der{\psi}{y} \right|_{y=\mT^I}
  +O(\ep^2).
\end{equation}  
Substituting it into (\ref{B-def}), we obtain
\begin{equation}
B=  \frac{b(x(\mT^I))}{J(\mT^I)}  +O(\ep^2).
\label{B-b1}
\end{equation}
This is nothing but the trapezoidal rule of the integral.
For later convenience, we also define
\begin{equation}
  R(y) \equiv \rho(x(y);V)
\label{R-def}
\end{equation}
for $y$ satisfying $y=T(x(y);V)$.


We here take $b(x)=p(T(x), \rho(x))$ and apply the above argument
to the identity for the pressure as
\begin{eqnarray}
\pex &=&  p(T(x),\rho(x)) \nonumber \\
     &=& \frac{1}{|I|} \int_{x_1}^{x_2} dx~ p(T(x),\rho(x)) \nonumber  \\
     &=&  \frac{p(\mT^I, R(\mT^I))}{J(\mT^I)}+O(\ep^2),
\end{eqnarray}
Since $p(\mT^I, \rho(x(\mT^I);V))=\pex$, we  obtain 
\begin{equation}
  J(\mT^I)=1+O(\ep^2),
\end{equation}  
and therefore (\ref{B-b1}) is simplified as
\begin{equation}
B=  b(x(\mT^I))  +O(\ep^2)
\label{B-b2}
\end{equation}
for any local thermodynamic quantities $b(x)$.
By setting $b(x)=\rho(x;V)$ in (\ref{B-b2}) and recalling (\ref{R-def}),
we have 
\begin{eqnarray}
\bar \rho^I &=&  \frac{1}{|I|} \int_{x_1}^{x_2} dx~ \rho(x;V) \nonumber  \\
          &=&  R(\mT^I)+O(\ep^2).
\label{bar-rho}
\end{eqnarray}
Finally, by applying (\ref{B-b2}) to
$b(x)=\phi(T(x;V),\rho(x;V))$, we obtain
\begin{eqnarray}
\Phi^I &=& \frac{V}{|I|} \int_{x_1}^{x_2}  dx~ \phi(T(x;V),\rho(x;V)) \nonumber \\
  &=& V \phi(\mT^I,R(\mT^I))+O(\ep^2) \nonumber  \\
  &=& V \phi(\mT^I,\bar \rho^I))+O(\ep^2).
\end{eqnarray}
For the last transformation we have used (\ref{bar-rho}).


\subsection{Derivation of (\ref{lam-1}) and (\ref{lam-2})} \label{lemma2}


We derive  (\ref{lam-1}) and (\ref{lam-2}) from the basic
assumptions on the steady state with an interface.  We 
express the steady state with an interface by 
$(\pex,N^\rmL_*,\Delta,N)$. As examples, we write 
\begin{eqnarray}
 \mT^\rmL{}_* &=&  \mT^\rmL(\pex, N^\rmL_*, \Delta, N), \\
 \mT^\rmG{}_* &=&  \mT^\rmG(\pex, N^\rmL_*, \Delta, N).
\end{eqnarray}
From the homogeneity in direction to perpendicular to $x$,
$ \mT^\rmL{}_*$ and  $\mT^\rmG{}_*$ are invariant for 
$(\pex,  N^\rmL_*, \Delta, N) \to 
(\pex, \lambda N^\rmL_*, \Delta,\lambda N)$.
We thus write 
\begin{eqnarray}
 \mT^\rmL{}_* &=&  \mT^\rmL(\pex, r_*, \Delta), \\
 \mT^\rmG{}_* &=&  \mT^\rmG(\pex, r_*, \Delta),
\end{eqnarray}
where
\begin{equation}
  r_*=\frac{N^\rmL_*}{N}.
\end{equation}  
By noting $\mT^\rmG{}_*-\mT^\rmL{}_*=\Delta/2$,
we express  $\mT^\rmL{}_* $ and $\mT^\rmG{}_* $ as 
\begin{eqnarray}
  \mT^\rmL{}_*
  &=& \Tc(\pex)-  \tau(\pex, r_*)\frac{\Delta}{2}+O(\Delta^2),
  \label{TmL}\\
  \mT^\rmG{}_*
  &=& \Tc(\pex)+ (1- \tau(\pex, r_*))\frac{\Delta}{2}+O(\Delta^2)
  \label{TmG}
\end{eqnarray}
for small $\Delta \ge 0$.


\begin{figure}[b]
\begin{center}
\includegraphics[scale=0.35]{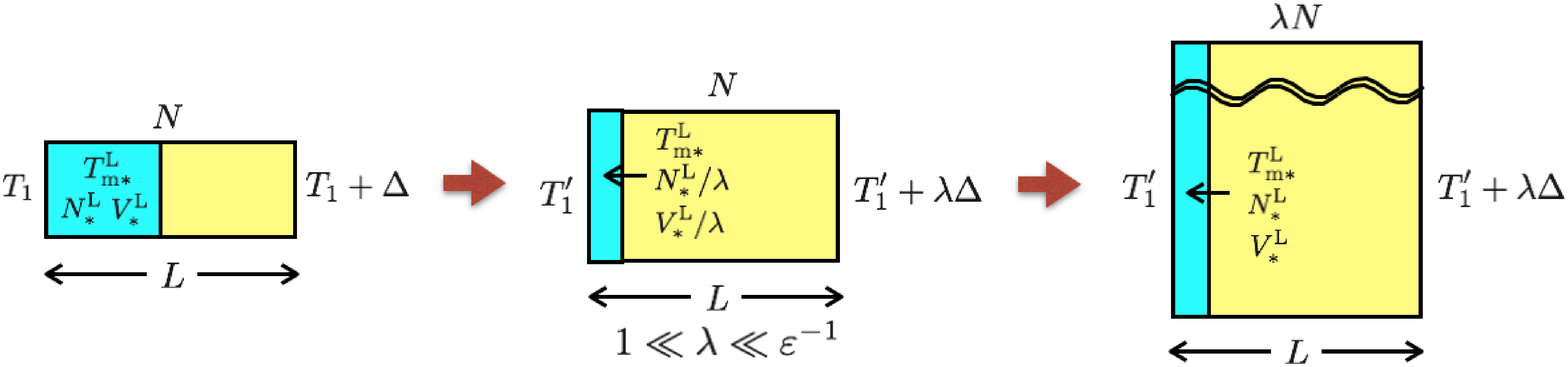}
\caption{}
\label{fig-s1}
\end{center}
\end{figure}

Now, we attempt to extend the gas region while keeping the thermodynamic
state in the liquid region.  See Fig. \ref{fig-s1}.
First, we increase $\Delta$ as
$\Delta \to \lambda \Delta$ with $\mT^\rmL{}_*$ fixed,
where we take $ \lambda  \ll \epsilon^{-1}$ such that the system
is still in the linear response regime. 
We estimate the $\lambda$
dependence of quantities in this asymptotic regime. First, 
the heat flux is proportional to $\lambda$ as a common value
to the liquid and the gas region. 
Since $\mT^\rmL{}_*$ is fixed, it is plausible that the temperature 
difference in the liquid region remains to be $O(\lambda^0)$,
and then  this leads that the horizontal length of the liquid 
region is proportional to $1/\lambda$. 
On the other hand, in the gas region, the temperature
difference is proportional to $\lambda$ and the horizontal length
is $L$ in the leading order estimate. Thus, the total volume
saturates at the finite value with which the whole system is occupied by
the gas with the temperature less than  $\Tc(\pex)+\lambda \Delta$. 
From these, we find that  the volume of the liquid region is
proportional to $1/\lambda$. Although its proportional coefficient
is not determined in the asymptotic region, we assume the scaling
relation that the volume of the liquid is  $V^\rmL_*/\lambda$ as it is 
consistent with the case $\lambda=1$.  Since $\pex$ and $\mT^\rmL{}_*$
is fixed in this operation, the density of the liquid
is also fixed. Thus, the particle number of the liquid becomes
$N^\rmL_*/\lambda$. Next, we increase $N$ as $N \to \lambda N$,
Then, we have $N^\rmL_*/\lambda  \to  N^\rmL_*$ and
$V^\rmL_*/\lambda  \to  V^\rmL_*$. That is, for a series of processes
\begin{equation}
  (\pex, N^\rmL_*, \Delta, N)
  \to
  (\pex, N^\rmL_*, \lambda \Delta, \lambda N),
\end{equation}
we have 
\begin{equation}
  \mT^{\rmL}  (\pex, N^\rmL_*, \lambda \Delta,  \lambda N)
= \mT^{\rmL}  (\pex,   N^\rmL_*, \Delta,  N).
\label{inv}
\end{equation}
Since this condition cannot be concluded from 
the local equilibrium thermodynamics, we impose (\ref{inv}) 
as a requirement for the steady state.

By using (\ref{TmL}) for (\ref{inv}), we have
\begin{equation}
\tau\left(\pex, \frac{r_*}{\lambda } \right)\lambda
=
\tau(\pex, r_*)
\label{inv3}
\end{equation}
for $1 \ll \lambda  \ll \epsilon^{-1}$.
Expanding 
\begin{equation}
\tau(\pex, r) = a_0+a_1 r +a_2 r^2+O(r^3) , 
\label{asympt}
\end{equation}
and substituting it to (\ref{inv3}), we obtain
\begin{equation}
(\lambda-1) a_0 +a_2 r_*^2(\lambda^{-1}-1)+\cdots=0
\label{asy2}
\end{equation}
for $1 \ll \lambda  \ll \epsilon^{-1}$. This gives
$a_0=a_2=a_n=0$ for $n \ge 3$. Thus,
\begin{equation}
\tau\left(\pex, r_* \right)=a_1 r_* ,
\label{result-1}
\end{equation}
which restricts the $r_*$ dependence of $\tau$.

Next, we  extend the liquid region while keeping the gas
region. In this case, we consider the  transformation 
$(\Delta, N) \to  (\lambda\Delta, \lambda N)$,
while fixing $(\mT^\rmG,N^\rmG_*)$. From (\ref{TmG}), we have
\begin{equation}
\left[
1-\tau\left(\pex, 1-\frac{N^\rmG_*}{\lambda N} \right)
\right] \lambda
=
1-\tau\left(\pex, 1-\frac{N^\rmG_*}{N} \right)
\label{inv-G}
\end{equation}
Expanding 
\begin{equation}
\tau(\pex, 1-u) = b_0+b_1 u +O(u^2) 
\label{asympt-gas}
\end{equation}
in the limit  $u \to 0$, we obtain
$b_0=1$ and $b_n=0$ for $n \ge 2$. 
This leads to 
\begin{equation}
b_1  \left( 1-r_* \right)
=
1-\tau\left(\pex, r_* \right).
\label{result-2}
\end{equation}
By combining  (\ref{result-1}) and (\ref{result-2}),
we obtain
\begin{equation}
b_1  ( 1-r_* )
=
1-a_1 r_*.
\label{result-3}
\end{equation}
Since this holds for any $r_*$, we obtain $b_1=a_1=1$.
By substituting (\ref{result-1}) with $a_1=1$ into 
(\ref{TmL}) and (\ref{TmG}),  we have arrived at
(\ref{lam-1}) and (\ref{lam-2}).


\section{A theory of pressure in the liquid-gas interface}


In steady heat conduction states under constant pressure $\pex$,
the temperature field $T(x)$ and the density field $\rho(x)$
satisfy
\begin{eqnarray}
  -\kappa(T(x),\rho(x))\partial_x T(x) &=& J, \label{eq-st-20}\\
  p(T(x),\rho(x))&=& \pex,  \label{eq-st-30}
\end{eqnarray}
where the heat flux $J$ is independent of $x$.
When $T_1 < \Tc(\pex) < T_2$, $T(x)$ and $\rho(x)$ for these
equations are not uniquely determined and they are
characterized by  the interface position  $\xc$.
Since the volume is determined from
\begin{equation}
  \frac{V}{L} \int_0^L dx~ \rho(x) = N \label{eq-st-40},
\end{equation}
we parametrize the solutions satisfying (\ref{eq-st-20}) and
(\ref{eq-st-30}) by $V$. In order to express this dependence
explicitly, in the main text, we use the notation
$T(x;V)$ and $\rho(x;V)$. They are characterized by 
\begin{eqnarray}
  -\kappa(T(x;V),\rho(x;V))\partial_x T(x;V) &=& J, \label{eq-st-2}\\
  p(T(x;V),\rho(x;V))&=& \pex,  \label{eq-st-3} \\
  \frac{V}{L} \int_0^L dx \rho(x;V) &=& N.  \label{eq-st-4}
\end{eqnarray}
These equations are equivalent to stationary solutions of
the standard hydrodynamic equation. As the position
of the interface is not determined in the framework of
standard hydrodynamics, it is reasonable to take the 
account of the higher-order derivative terms
in the hydrodynamic equation. 
In this section, we consider the pressure associated with
the density gradient in the interface, which gives  higher
order terms in the hydrodynamic equation. We show that
the steady-state value of $V$, which is denoted by $V_*$
in the main text, becomes accessible by adopting this extension.


In the standard hydrodynamics, the interface
is treated as the discontinuous surface of the density field.
Here, we treat the interface
as a transition layer of the width $\xi$ and  consider
the density profile in the layer. 
Since the density gradient in the transition layer  is much larger
than that in the bulk, we expect that its contribution to the
pressure is comparable with $p(T,\rho)$. Concretely, 
noting that $\rho(L-x;V)$ and $T(L-x;V)$ satisfy the same equation
for $\rho(x;V)$ and $T(x;V)$ when reversing $T_1$ and $T_2$,
we consider the lowest order form of  gradient expansion of the
pressure balance equation.  We then find that  (\ref{eq-st-3}) is
replaced by 
\begin{equation}
p(T(x;V),\rho(x;V))+
\frac{1}{2}d_1(\rho)\xi^2(\partial_x \rho)^2
  -d_2(\rho) \xi^2 \partial_x^2 \rho  +d_3(\rho)J \xi \partial_x \rho
= \pex ,
\label{eq-target}
\end{equation}
where $d_1$, $d_2$  and $d_3$ are some functions of $\rho$
which are assumed to be independent of $\xi$. More precisely,
$d_1$ and $d_2$ are functions of $(T,\rho)$ whose functional
forms are specific to the type of material, while the functional
form of $d_3$ may be influenced by the system condition
$(\pex, T_1, T_2)$. The terms proportional to $d_1$ and $d_2$ were
proposed by van der Waals and adopted in generalized hydrodynamic
equations referred in the main text as [9,10].
The term proportional to $d_3$ does not appear in these approaches.
The density gradient terms are relevant only in the transition layer
where $|\xi \partial_x\rho| \simeq |\rho^\mathrm{L}-\rho^\mathrm{G}|$, 
whereas $|\xi \partial_x\rho|\simeq \rho \xi/L \ll\rho^\rmG$ for its outside.
$d_2(\rho) >0 $ for the stability of the density profile. 
Furthermore, since the temperature $T$ is continuous at $x=\xc$ and $\xi$ 
is sufficiently small, the temperature $T(x)$ does not vary so much in the 
transition layer. Therefore,  temperatures inside the transition layer 
may be evaluated as $T(X)$ for thermodynamic quantities.


Equilibrium cases are described by (\ref{eq-st-2}), (\ref{eq-st-4}),
and (\ref{eq-target}) with  $J=0$. Then, following the argument
by van der Waals, we derive the relation 
\begin{equation}
  d_1(\rho)= 2 \frac{d_2(\rho)}{\rho}- d_2'(\rho),
\label{d-c}
\end{equation}
where the prime represents the derivative with respect to $\rho$.
See Sec. \ref{vw-relation} for the derivation. We assume that
this relation also holds in the linear response regime. 


Now, we proceed to nonequilibrium cases.
We show that there is the unique solution $\rho(x;V_*)$ that satisfies
(\ref{eq-st-2}), (\ref{eq-st-4}), and (\ref{eq-target}).
This solution also  determines the value of $\xc_*$. Concretely,
we define $\mu(T,\rho)$ as
\begin{equation}
  \mu(T,\rho) \equiv \frac{f(T,\rho)+p(T,\rho)}{\rho},
\end{equation}
where $f(T,\rho)=F_{\rm vW}(T,V,N)/V$ with $\rho=N/V$.
We then find that $\xc_*$ satisfies 
\begin{equation}
\lim_{\xi/L \to 0}
\left[ \mu(T(\xc_*;V_*),\rho(\xc_*+K \xi;V_*)) 
  -\mu(T(\xc_*;V_*),\rho(\xc_*-K \xi;V_*))
\right] = J \Delta_{\rm gap},  
\label{mu-jump}
\end{equation}
which is  derived in Sec. \ref{mu-gap},
where a numerical constant $K \gg 1$ is fixed.
Here, $\Delta_{\rm gap}$
takes a finite value in the limit $\xi/L \to 0$. 
In equilibrium cases $J =0$, (\ref{mu-jump})
means the continuity of $\mu(T(x;V_*),\rho(x;V_*))$ at $x=\xc_*$.
This balance relation determines the value of $V_*$, which
is consistent with the variational principle in equilibrium thermodynamics.
However, in heat conduction states, $\mu(T(x;V_*),\rho(x;V_*))$
is discontinuous at $x=\xc_*$. This implies that super-cooled
gas or super-heated liquid becomes stable
as a local equilibrium state in the heat conduction.
This is consistent with the prediction in the main text.

\subsection{Proof of (\ref{mu-jump})}\label{mu-gap}

We first define
\begin{align}
{\cal M}(T,\rho;\pex)&=\frac{f(T,\rho)+\pex}{\rho}.
\end{align}
Using the relation 
\begin{equation}
 p(T,\rho) =-\pder{ \left(f(T,\rho)\rho^{-1}\right)}{\rho^{-1}},
\label{p-f}
\end{equation}
we express $p-\pex$ as
\begin{equation}
  p(T,\rho)-\pex =\rho^2\pder{{\cal M}}{\rho}.
\label{p-pex}
\end{equation}
(\ref{eq-target}) is then rewritten as
\begin{equation}
\rho^2\pder{{\cal M}}{\rho}+\frac{d_1(\rho)}{2}\xi^2(\partial_x \rho)^2
  -d_2(\rho) \xi^2\partial_x^2 \rho  +d_3(\rho)J \xi\partial_x \rho
  =0 .
  \label{eq-target2}
\end{equation}
We further note 
\begin{eqnarray}
\pder{}{x}\left(\frac{d_2(\rho)}{\rho^2}(\partial_x\rho)^2\right)
=
-\frac{2\partial_x\rho}{\rho^2}\left[
\frac{d_1(\rho)}{2}(\partial_x\rho)^2-d_2(\rho)\partial_{x}^2\rho\right],
\label{identity}
\end{eqnarray}
where we have used (\ref{d-c}) in order to remove $d_2'(\rho)$.
Since (\ref{eq-target2}) is transformed into
\begin{equation}
\pder{{\cal M}}{\rho}+\frac{d_3(\rho)}{\rho^2} J \xi\partial_x \rho
=-\frac{\xi^2}{\rho^2}
\left[\frac{d_1(\rho)}{2}(\partial_x \rho)^2
  -d_2(\rho) \partial_x^2 \rho  \right]
  ,
  \label{eq-target3}
\end{equation}
(\ref{identity}) and  (\ref{eq-target3}) lead to
\begin{eqnarray}
\pder{}{x}\left({\cal M}-\frac{d_2(\rho)}{2\rho^2}\xi^2(\partial_x\rho)^2\right)=
-\frac{d_3(\rho)}{\rho^2}J\xi(\partial_x\rho)^2, 
\label{balance}
\end{eqnarray}
where it should be noted that
$\partial_x{\cal M}=(\partial_{\rho}{\cal M})(\partial_x\rho)$.

Now, we integrate (\ref{balance}) over the transition layer 
at $x=\xc_*$. 
(\ref{balance}) leads to
\begin{equation}
  \left. \left[
 {\cal M} -\frac{d_2(\rho)}{2}\xi^2\left(\frac{\partial_x \rho}{\rho} \right)^2
    \right] \right|^{\xc_*+K \xi}_{\xc_*-K \xi}=
 -J \int_{\xc_*-K \xi}^{\xc_*+K \xi} dx~ d_3(\rho)\xi\left(\frac{\partial_x\rho}{\rho}\right)^2 ,
\label{balance2}
\end{equation}
where  $K$ is a large constant.
Since $\partial_x\log \rho=O(1/L)$ and ${\cal M} =\mu$ at $x=X_*\pm K\xi$,
which is outside of the transition layer, (\ref{balance2}) is simplified as
\begin{equation}
   \mu(T(X_*;V_*),\rho(X_*+K \xi;V_*))-\mu(T(X_*;V_*),\rho(X_*-K \xi;V_*))
   =-J \int_{\xc_*-K \xi}^{\xc_*+K \xi}
   dx~ d_3(\rho)\xi\left(\frac{\partial_x\rho}{\rho}\right)^2
   +O\left(\frac{\xi^2}{L^2} \right),
\end{equation}   
where we have used the continuity of temperature in the transition layer.
By setting  $\rho(x)=\rho(X_*+\xi t)$ 
as  a function of $t$, we write
\begin{equation}
  \int_{\xc_*-K \xi}^{\xc_*+K \xi}dx~
  d_3(\rho)\xi\left(\frac{\partial_x\rho}{\rho}\right)^2
=  \int_{-K}^{K}
dt~ d_3(\rho)\left( \partial_t \log \rho \right)^2 .
\end{equation}
Thus, we reach (\ref{mu-jump}), where
\begin{equation}
  \Delta_{\rm gap}=
-\int_{-\infty}^{\infty}
dt~ d_3(\rho)\left( \partial_t \log \rho \right)^2,
\label{Delta-gap}
\end{equation}
which is independent of $\xi$. 
(\ref{Delta-gap}) indicates that $\Delta_{\rm gap}$ is finite in general.

\subsection{Proof of (\ref{d-c})}\label{vw-relation}

We consider  equilibrium states at which liquid and
gas coexist in a container with constant volume. 
For simplicity, we assume that the density depends
only on $x$. We determine the density profile for
this system by assuming the free energy 
\begin{equation}
  F( [\rho])
  = \frac{V}{L}
  \int_0^L dx \left[f(T,\rho(x))+ \frac{c(\rho(x))}{2}
    \xi^2 (\partial_x  \rho)^2  \right],
  \label{Fdef}
\end{equation}
where the term
$ c(\rho)\xi^2(\partial_x  \rho)^2 /2$  represents the effective
energy density associated with the density gradient. This term may be
relevant at the interface where the density is discontinuous
in the thermodynamic description. We apply the variational 
principle for  $ F( [\rho])$ under the condition
\begin{equation}
\frac{V}{L} \int_0^L dx \rho(x)=1.
\label{const}
\end{equation}
Noting that
\begin{equation}
\mu(T,\rho)=\pder{f}\rho,
\end{equation}
we obtain the variational equation for (\ref{Fdef}) as
\begin{equation}
  \mu(T,\rho)+\frac{c'(\rho)}{2} \xi^2 (\partial_x \rho)^2
  - \xi^2 \partial_x( c(\rho)  \partial_x \rho)+\lambda=0,
\label{var-eq}
\end{equation}
where $\lambda$ is the Lagrange multiplier whose value
is determined by (\ref{const}).

Next, we derive the pressure balance equation, which
provides the equation of state for the system with the interface.
By substituting 
\begin{equation}
  \mu(T,\rho)=\frac{f(T,\rho)+p(T,\rho)}{\rho}
\end{equation}  
into (\ref{var-eq}) and
taking the differentiation in $x$, we obtain
\begin{align}
  \partial_x\left( \frac{f+p}{\rho} \right)
  + \xi^2\partial_x\left[ \frac{c'(\rho)}{2} (\partial_x \rho)^2 \right]
  -   \xi^2 \partial_x^2(c(\rho) \partial_x \rho) = 0.
\label{s27}
\end{align}
Since $\partial_x f=\partial_\rho f\partial_x \rho$, we have an equality
\begin{align}
  \partial_x\left(\frac{f+p}{\rho} \right) &= \rho^{-1}
  \left(
  \partial_\rho f \partial_x\rho+\partial_x p -\frac{f+p}{\rho}\partial_x\rho
  \right)\nonumber\\
  &=\frac{\partial_x p}{\rho}.
\label{s28}
\end{align}  
From (\ref{s27}) and (\ref{s28}), we obtain 
\begin{equation}
  \partial_x  p
  + \rho \partial_x\left[ \frac{c'(\rho)}{2} \xi^2 (\partial_x \rho)^2 \right]
  -  \rho  \xi^2 \partial_x^2(c(\rho) \partial_x \rho) = 0.
\end{equation}
Here, from the elementary calculation, we can show
\begin{equation}
  \partial_x  p + \partial_x
  \left[ \frac{1}{2}(c(\rho)-\rho c'(\rho))\xi^2(\partial_x \rho)^2
    -c(\rho) \rho \xi^2 \partial_x^2 \rho  \right]=0,
\end{equation}
which leads to
\begin{equation}
  p(T,\rho)+ 
  \left[\frac{1}{2}d_1(\rho)\xi^2(\partial_x \rho)^2-d_2(\rho)\xi^2 \partial_x^2 \rho
    \right]=\pex,
\label{ex-vW}  
\end{equation}  
where
\begin{eqnarray}
  d_1(\rho) &=&  c(\rho)-\rho c'(\rho), \\
  d_2(\rho) &=&  c(\rho)\rho .
\end{eqnarray}
From these, we obtain (\ref{d-c}).


\begin{thebibliography}{99}
\bibitem{fn:GLTC}
  When both the temperatures of the baths are larger (or smaller)
  than $\Tc(\pex)$, the system is filled with the gas (or liquid).
\bibitem{fn:hydro}  
  The stationary solutions of the standard hydrodynamic equations
  are obtained from (\ref{p-balance}), (\ref{Fourier}), and 
  (\ref{v-det}). These are parametrized by the interface position.
\bibitem{PA}
  A solution is uniquely selected, for example,
if the interface temperature 
  is equal to $\Tc(\pex)$, which is equivalent to the application
  of the equilibrium van der Waals theory \cite{vW} to 
  heat conduction states \cite{Bedeaux03,Onuki}. However, the validity
  of this assumption is not established  yet. See also the discussion in the
  paragraph involving (\ref{int-pressure}).
\bibitem{fn:vW}  The van der Waals free energy $F_{\rm vW}(T,V,N)$
  is not a thermodynamic free energy. Indeed, it is not a convex
  function in $V$. Instead, $F_{\rm vW}(T,V,N)$ may be interpreted 
  as a mesoscopic free energy
  obtained by integrating microscopic degrees of freedom in 
  ranges less than the nucleation size.
\bibitem{fn:Legendre}
The result $V_*(T,\pex)$ is equivalent to that obtained by 
the Maxwell construction (the equal area law) from the van 
der Waals equation of state (\ref{van-der}).
\bibitem{fn:gen}
Even for general cases with a short range
interaction, we may employ  experimental data
for the equation of states $p=p^\rmL(T,\rho)$
and $p=p^\rmG(T,\rho)$ for liquids and gases, respectively.
Note that these may be applied even to metastable states. 
Then, from the experimental data, a mesoscopic
free energy $F_{\rm vW}(T,V)$ is obtained in a certain range of $V$, which is
sufficient to apply our theory to general cases.
\bibitem{fn:bubbles}
We assume that bubbles or droplets are not observed.
\bibitem{fn:ss-ch}
The thermodynamic state in the liquid region is described by
the pressure, the mean temperature and the particle number 
of the liquid, as shown from (\ref{Phi-def-main}).
\bibitem{fn:gT}
 For $(T_1,T_2)$, we calculate (\ref{vf}) and (\ref{global-T}), which yields the value of ${\cal G}(V; \bT,\pex,\Delta)$.
\bibitem{fn:LM}
  In the conventional variational principle of equilibrium cases,
  the pressure of the system is not fixed to $\pex$.  The variational
  function is then found to be minimum when $p=\pex$. 
  To quickly identify the minimum, we consider cases $p=\pex$ from the outset.
\bibitem{fn:Tm}
We can prove this relation by using (\ref{Phi-def-main}).
The higher order estimation is also obtained by the generalization
of the trapezoidal rule for the approximation of integrals.
\bibitem{NS}
We can show 
$dG=-S_* d \tilde T+V_* d \pex$ for each $\Delta$,
where $S_*$ is the spatial integration of 
the local entropy density.
Details will be reported in another paper.
\end{thebibliography}
\end{document}